\documentclass[prl,aps,twocolumn,showpacs]{revtex4}
\usepackage{graphicx}
\usepackage{dcolumn}
\usepackage{bm}
\usepackage{amsmath}
\usepackage{amssymb}
\usepackage{amsfonts}
\usepackage{times}

\begin{document}
\title{Forced tearing of ductile and brittle thin sheets}
\author{T. Tallinen$^1$}
\author{L. Mahadevan$^1$}
\affiliation{School of Engineering and Applied Sciences, Harvard University, Cambridge, MA 02138, USA}

\date{\today}
\begin{abstract}
Tearing a thin sheet by forcing a rigid object through it leads to complex crack morphologies; a single oscillatory crack arises when a tool is driven laterally through a brittle sheet, while 
two diverging cracks and a series of concertina-like folds forms when a tool is forced laterally through a ductile sheet. On the other hand, forcing an object perpendicularly through the sheet leads to radial petal-like tears in both ductile and brittle materials. To understand these different regimes we use a combination of experiments, simulations and simple theories.  In particular,  we describe the transition from brittle oscillatory tearing via  a single crack to ductile concertina tearing  with two tears by deriving laws that describe the crack paths and wavelength of the concertina folds and provide a simple phase diagram for the morphologies in terms of the material properties of the sheet and the relative size of the tool.
 
\end{abstract}
\maketitle

The failure of thin sheets by fracturing, tearing and peeling occurs naturally on a range of scales - from the everyday world of opening an envelope or other package  to the grounding of ships \cite{atkins}, from the failure of armor plating to the cracking of geological structures \cite{hoppa}.  The study of the fracture and tearing of thin sheets is challenging as it couples the geometry of large out-of-plane deformations to the failure of the material. Deformations leading to in-plane stretching, for example because of constraints in the far field, play a crucial role in determining the resulting complex crack morphologies. While  early work focused  primarily on the role of plastic deformations \cite{atkins, wierzbicki} there has been a recent surge of interest  \cite{ghatak, roman,audoly, atkinsw, bayart, cohen} in understanding these questions quantitatively in brittle elastic materials. This is driven in large part by the focus on understanding the mechanical behavior of thin films at the meso/nano scale, in developing of assays for the measurement of material properties and creating new functional structures \cite{hamm, sen}.   

\begin{figure}[t]
\begin{center}
\includegraphics[width = 85mm]{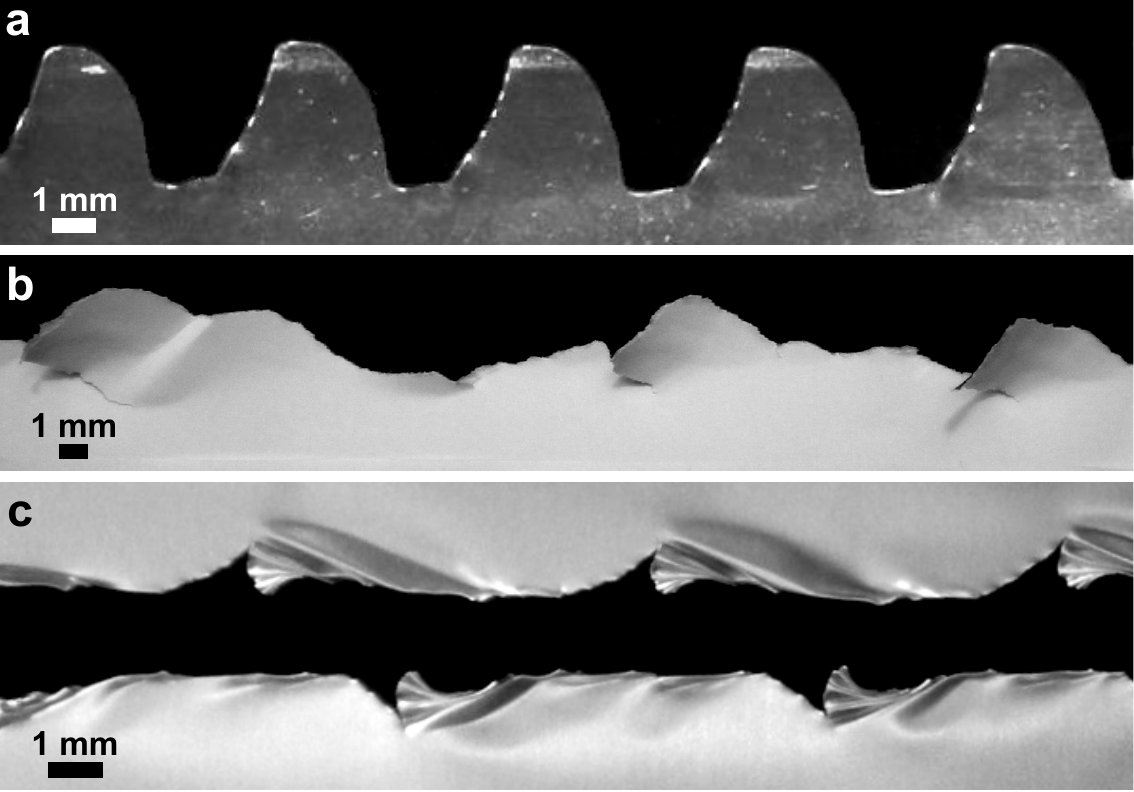}
\end{center}
\caption{Complex morphologies of cracks in a sheet of (a) brittle polymeric material, (b) paper, and (c) aluminum film torn by a cylindrical tool that moves in the plane of the sheet.
}
\label{fig1}
\end{figure}

An important class of fracture and failure in thin sheets is that due to the forcing of a solid object or tool through it, either in the plane of the sheet or transverse to it. Forced tearing of a relatively brittle sheet, made of acetate, by the motion of a tool  in the lateral direction leads to oscillatory fracture (Fig. \ref{fig1}a) if the tool diameter $D$ is large in comparison to the sheet thickness $h$. For this system, various levels of approximation yield a hierarchy of models for crack paths \cite{ghatak, roman, audoly, atkinsw}, all of which show that the amplitude and wavelength of the oscillation scale with $D$, consistent with experimental observations. Furthermore the dynamics of the single crack in this case shows a characteristic stick-slip behavior associated with the transitions between bend-dominated and stretch-limited motion of the tool. Given the strongly geometric flavor of all models for this phenomena, a natural question is the following:  how would a ductile sheet that can deform plastically before fracture behave when forced similarly? Here we attempt to synthetically understand the phase space of the different crack morphologies that arise in the tearing of a thin sheet as a function of the type of forcing, the geometry of the sheet and its material properties.

In Fig. \ref{fig1}b we show the results of fracture in a thin sheet of paper  driven by the lateral motion of a tool;  paper is more ductile than acetate, and this leads to disordered version of the oscillatory fracture seen in acetate sheets. In Fig. \ref{fig1}c we show the results of a similar experiment carried out with an aluminum sheet, which is far more ductile, and see two marked differences:  the sheet fractures by forming two cracks, and a series of irreversible periodic folds forms along the  crack edges.  If the same experiment is now carried out in the presence of a rigid substrate that supports the thin aluminum sheet using a tool that leans forward in the direction of motion, we see a beautiful 'concertina' tearing pattern (Fig. \ref{fig3}a)  first studied in the context of the failure of relatively thick metallic plates \cite{wierzbicki}, and understood using a plastic kinematic deformation model. For very thin sheets or for materials with a high yield stress, however, this  analysis is not valid since the deformations ahead of the crack tips are predominantly elastic.  

To understand the role of ductility in thin sheet fracture, we start with numerical simulations that combine finite elastic-plastic deformation with a simple fracture criterion to synthesize the variety of observed fracture morphologies in brittle and ductile sheets. Our simulations are based on a discrete-element model that consists of mass points arranged in a random lattice with average area $h^2$ per site, where $h$ is the thickness of the sheet \cite{sm}. Lattice sites have translational and rotational degrees of freedom, and each pair of neighboring  points is connected by elastic-plastic element with bending, shear and tensile stiffness, similar to one used to study crumpling \cite{tallinen} and brittle fragmentation \cite{astrom}. Lattice sites have six neighbors on average. We assume an ideal elastic-plastic stress-strain relation with Young's modulus $E$ and yield stress $\sigma_y$. Elements are forced to break when their tensile or shear strain exceeds $\gamma_s$. For ductile materials $\sigma_y/E < \gamma_s$, i.e. the material deforms plastically before fracture. For $\sigma_y/E \ll \gamma_s$ we obtain the toughness $\Gamma \sim \sigma_y h$ typical for thin metal sheets \cite{hutchinson}, while in brittle case $\Gamma \approx Eh\gamma_s^2$ \cite{sm}. In our simulations $\Gamma$ is the sum of energy of broken elements and plastic dissipation per unit area of crack extension. The above estimates for $\Gamma$ provide a convenient way to relate our simulations to experiments; however the connection of $\gamma_s$ to real fracture strains is more complicated. Self-avoidance of the sheet is generated via an elastic repulsion with range $h$ between any two non-neighboring lattice sites. The dynamics of the system is simulated by solving Newton's equations of motion. We start with a long sheet with dimensions 320$h$ $\times$ 1280$h$ that is clamped along its lateral edges with no pre-strain and a small notch is introduced on one of the short edges via which a tool is introduced to cut the sheet. The tool is moved with a given velocity that is small enough that the inertia is negligible. We model the interaction between the tool and the sheet using a simple Coulombic frictional law, with a coefficient $\mu$ for both static and dynamic friction.

\begin{figure}[t]
\begin{center}
\includegraphics[width = 85mm]{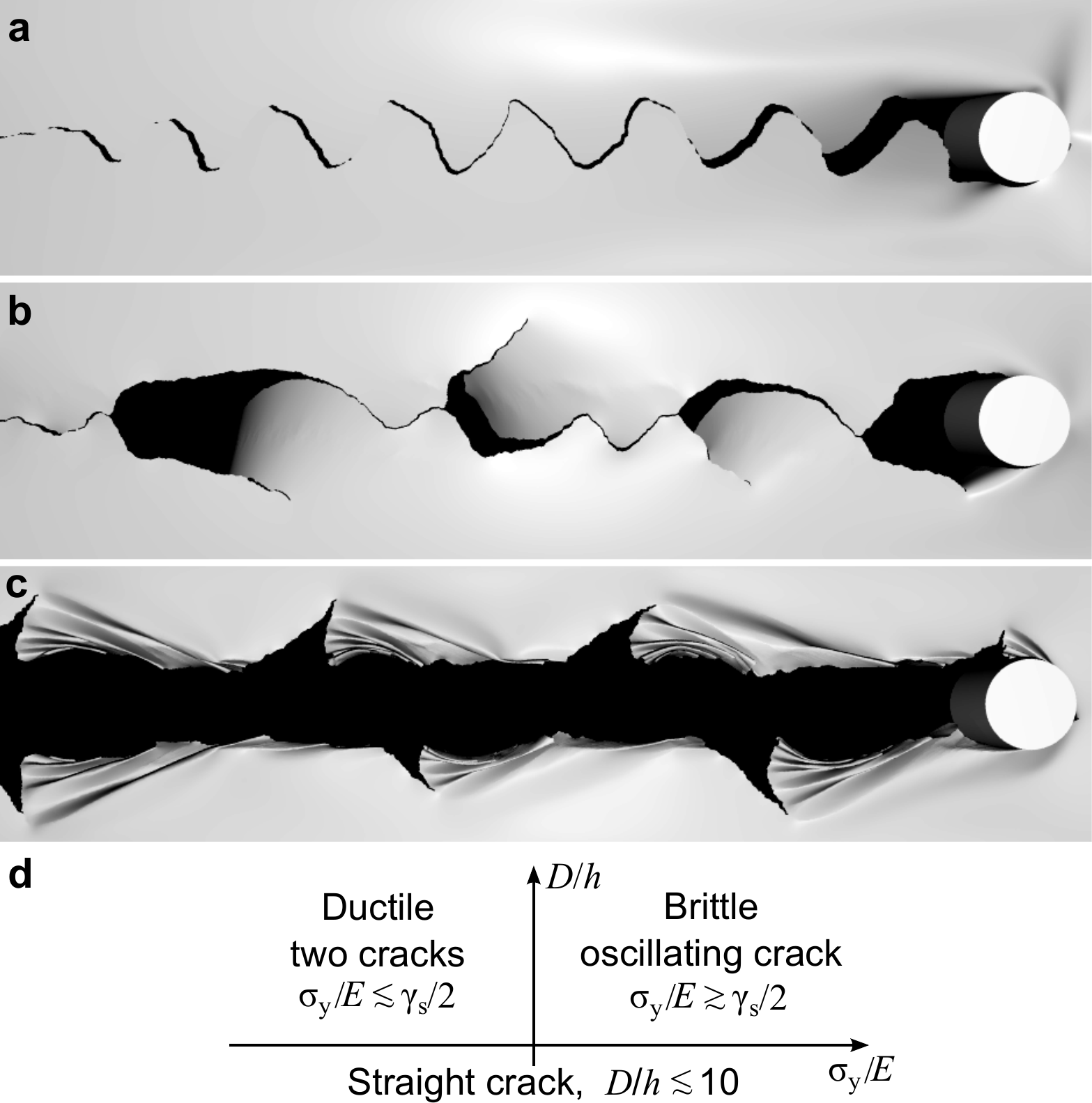}
\end{center}
\caption{Simulations that reproduce the experiments shown in Fig. 1.
(a) An oscillating crack in a  brittle sheet.
(b) Disordered tongues  in a weakly ductile sheet with $\sigma_y/E = \gamma_s/4$.
(c) Periodic concertina tearing in ductile sheet with $\sigma_y/E = \gamma_s/40$.
Contact between the tool and sheet has a friction coefficient $\mu = 0.25$ and
tool diameter $D = 100h$.
(d) Phase diagram of tearing morphologies as a function of tool size and material behavior characterized by the relative magnitude of the yield strain $\sigma_y/E$ to the fracture strain $\gamma_s$.
 }
\label{fig2}
\end{figure}

\begin{figure*}
\includegraphics[width = 150mm]{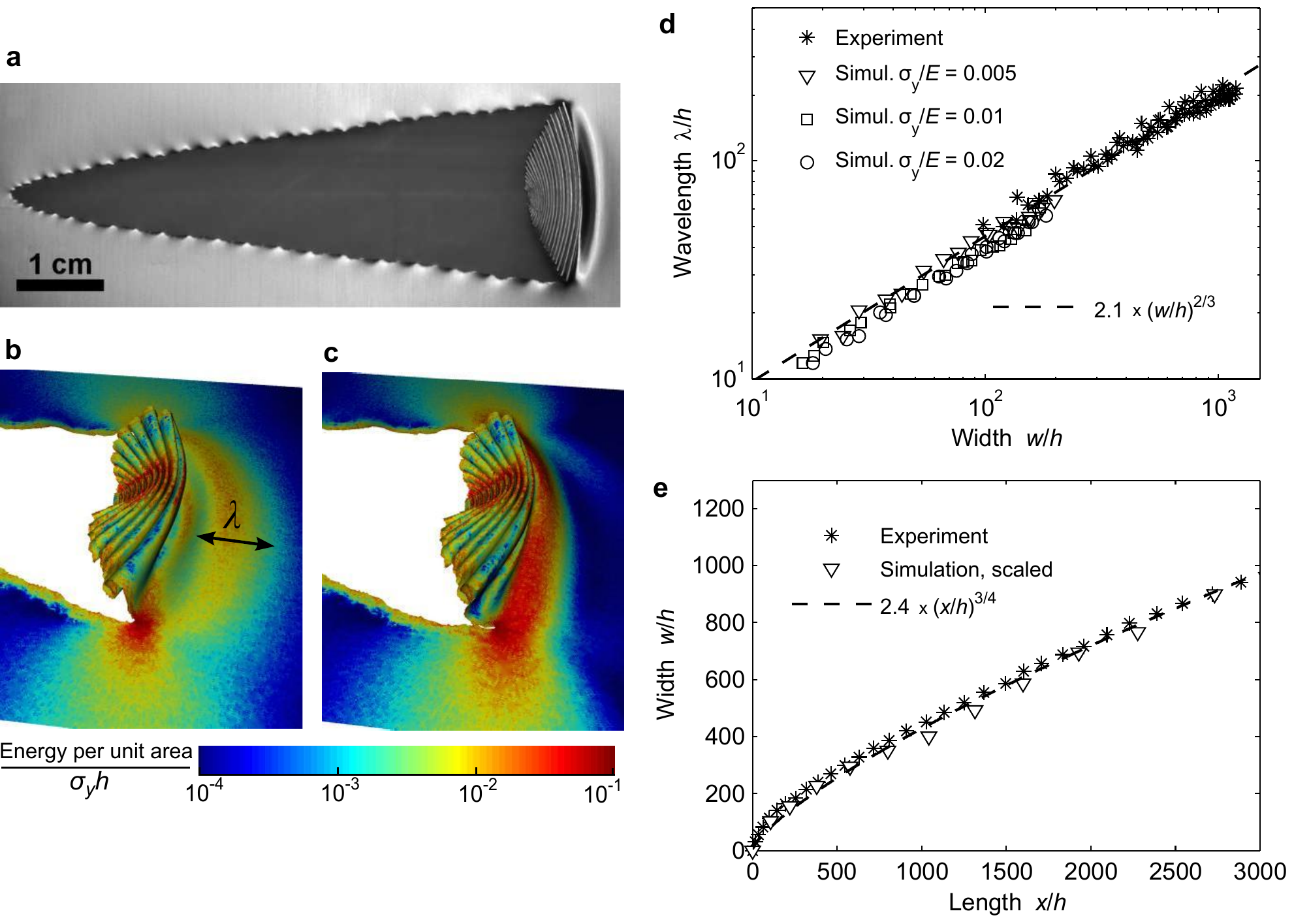}
\caption{(color online). Concertina tearing of (a) aluminum film and (b,c) simulated ductile sheet on a rigid substrate. In (b) the process starts with the formation of a wrinkle of wavelength $\lambda$ that is then stretched (c) before the cracks propagate.  Color indicates elastic stretching energy density (red = high, blue = low).
(d) Experimental and simulated wavelength of concertina folds are plotted as a function of width $w$ of the tongue and compared with $\lambda/h \sim (w/h)^{2/3}$. 
Each plot includes data from three (experiments) or two (simulations) independent measurements. In each simulated case $\sigma_y/E = \gamma_s/20$.
(e) Tongue width $w$ is plotted as a function of length $x$ of the tongue and compared with $w/h \sim (x/h)^{3/4}$. 
Simulated $x$ and $w$ are magnified by factors $8$ and $8^{3/4}$, respectively, for better visual comparison. }
\label{fig3}
\end{figure*}

We first replicate numerically the experiments on cutting a sheet with a cylindrical tool of diameter $D$ that ploughs through a sheet of thickness $h$  ($D \gg h$) laterally \cite{ghatak, roman,sm}. For a brittle sheet we observe an oscillating crack path as shown in Fig. \ref{fig2}a. Following an initial transient the motion of the crack settles into that observed in experiments (Fig. \ref{fig1}a). Furthermore, the cutting force \cite{sm} is highly oscillatory as elastic energy is stored and released in dynamic bursts of fracture \cite{ghatak, audoly}. We confirm that fracture occurs by in-plane stretching when the crack tip is at either extreme laterally and by out-of-plane shear when the tip is close to the center-line \cite{ghatak}. We use $\gamma_s = 0.4$ to obtain $\Gamma/Eh \sim 10^{-1}$, a value typical for polymeric sheets used previously \cite{ghatak, audoly}. The relative toughness here is much higher than that of ceramic materials and glass,  where forced failure occurs through brittle fragmentation.

Keeping the geometry of the sheet and tool the same, we now vary the onset of plastic yielding by reducing the yield strain $\sigma_y/E$ so that it is smaller than the fracture strain $\gamma_s$. Oscillatory motion of a single crack persists till $\sigma_y/E \approx \gamma_s/2$; as the yield strain is reduced further, the tool propagates by forming two parallel cracks rather than a single oscillatory one. Our simulations indicate that this transition is independent of $D$ when $D \gg h$. However, for $D \sim h$ stable straight cracks have been reported for both brittle \cite{ghatak} and ductile \cite{atkins} sheets. While the discrete nature of the lattice limits accuracy in this limit, we find that for  $D \lesssim 10h$ crack paths are also straight in both the brittle and ductile case.

For $\sigma_y/E \lesssim \gamma_s/2$ and $D \gg h$  the nucleation of new cracks arises when a single crack is driven away from the middle-line by the advancing tool. In a brittle sheet,  this is followed by a dynamical burst returning the crack tip back to the middle. In ductile sheet, however,  this step is hindered by plastic deformation and a new crack nucleates at the point diametrically opposite where the crack tip meets the tool. Then, both cracks propagate simultaneously while extruding a tongue of material ahead of the tool. Whether or not this tongue folds up into a concertina pattern depends on the ductility of the material. For weakly ductile materials, the tongue bends out of the plane without forming permanent folds (Fig. \ref{fig2}b) resulting in fracture patterns seen in torn paper (Fig. \ref{fig1}b). For strongly ductile materials, with $\sigma_y/E \ll \gamma_s$ these tongues appear on alternating sides of the tool in a form of short concertina folds (Fig. \ref{fig2}c), as  seen in torn aluminum films (Fig. \ref{fig1}c). The size of the folded regions scales with $D$ and increases with increasing friction between the tool and the sheet. Finally, by introducing a pair of initial cracks on both sides of the tool or a rectangular tool, we find that a stable series of concertina folds results. Thus, although the morphology of tearing of thin sheets is dependent on tool shape, contact friction and the  mechanism of crack initiation, at a primitive, coarse level the patterns are characterized by material behavior; ductile tearing is characterized by the formation of a tongue between two cracks that may fold up ahead of the tool,  while brittle tearing is characterized by the oscillatory motion of a single crack. This simple phase diagram is summarized in Fig. \ref{fig2}d in terms of the tool size and material behavior of the sheet.  

To analyze concertina tearing in more detail, we numerically simulate the configuration of Fig. \ref{fig3}a by including the effect of a solid substrate below the sheet using a simple repulsive potential. Tearing is driven by a rectangular tool of width 15$h$ that leans into the direction of motion at an angle of 45$^{\circ}$, just as in the experiment: this leads to stable concertina tearing as the tongue is confined between the tool and substrate. When the tool advances, we observe new folds forming in two steps: 1) the sheet ahead of the crack tip wrinkles, and 2) the crack tips advance by passing on either side of the wrinkle just formed, which now folds plastically and runs into folds formed in earlier cycles (Figs. \ref{fig3}b,c). The resisting force \cite{sm} in a cycle of folding and tearing increases as wrinkles form, reaches a peak just before crack advances due to strong stretching, and then falls before the  cycle starts anew; unlike in brittle tearing, fracture is quasistatic and tensile. This form of tearing is also insensitive on the form of tool used.

To determine  the wavelength of the folds as well as the crack paths, we must consider the stress field ahead of the two crack tips as a function of tool motion. Once a stack of folds has formed, the force due to tool motion is transmitted through  these to the sheet ahead over a width $w$, which deforms out of the plane with amplitude $A$ and wavelength $\lambda$, measured in the direction of tool motion. The resulting stretching strain $(A/w)^2$ is primarily in the direction perpendicular to tool motion. The stretching energy in an area $\lambda w$ scales as $U_S \approx Eh (A/w)^4 \lambda w$ while the bending energy in the same area $U_B \approx Eh^3(A/\lambda^2)^2 \lambda w$.  Minimization of $U_S + U_B$ yields $\lambda \approx w \sqrt{h/A}$ \cite{cerda}. The wavelength, however, is limited by $\lambda \gtrsim 2A$ which leads to an expression for wrinkle wavelength given by
\begin{equation} \label{lambda}
\lambda = c_{\lambda}w^{2/3}h^{1/3}.
\end{equation}
Further compression only folds the wrinkles plastically. To test the results of our theory and numerical simulations, we carried out tearing experiments using  aluminum film of thickness $h = 24$ $\mu$m, yield stress $\sigma_y \approx 0.003E$ and toughness $\Gamma \approx \sigma_yh$ \cite{hutchinson}. Using image analysis, we extracted  the wavelength $\lambda$ of the folds as a function of the width $w$ of the tongue and compared them to the results of simulations (Fig. \ref{fig3}d), and find an excellent match between simulation, theory and experiment, even including the weak dependence of $\lambda$ on the yield stress.  Both experimental and simulation data are well described by equation (\ref{lambda}) with the constant $c_{\lambda} = 2.1 \pm 0.1$ even for materials with a high yield stress when deformation ahead of the crack tips is primarily elastic. Interestingly, our elastic analysis for $\lambda$ leads to an expression that is identical to one based on a purely plastic analysis \cite{atkins, wierzbicki}, despite the different material assumptions, owing as is typical in these situations to the strong constraints imposed by the geometry of the system. For very thin sheets, when $h/w \rightarrow 0$ our elastic analysis is always valid for since the maximum strain of bending during the wrinkle formation becomes vanishingly small compared to the yield strain.

To understand the crack paths  we assume, as is usual, that crack extension occurs when the energy available for crack growth is sufficient to overcome
the resistance of the material \cite{audoly, anderson}. For our system this  leads to the expressions 
\begin{align}
\frac{\partial U}{\partial x}&=2\Gamma h \cos\theta \\
\frac{\partial U}{\partial w}&=-2\Gamma h \sin\theta,
\end{align}
where equations (2) and (3) correspond to force balance parallel and perpendicular to the middle line, respectively, $U$ is the elastic energy, and $\theta$ is the angle of crack paths with the middle line. The factors of two follow from both cracks advancing simultaneously. Dividing (3) by (2) we obtain $\tan\theta = dw/dx =-\frac{\partial U}{\partial w}/\frac{\partial U}{ \partial x} $. For high $w/h$ the energy of the wrinkle formation regime is insufficient for propagation of the cracks; stronger in-plane stretching is required so that the whole fold gets stretched (see Figs. \ref{fig3}a,b for a wrinkle forming and stretching to the point of fracture). The energy $U$ thus has a form similar to $U_s$ given above. Since that energy is distributed on an area of width $\lambda$ ahead of the crack tips, $\frac{\partial U}{\partial x} \approx U/\lambda$, while $\frac{\partial U}{\partial w} \sim -U/w$ so that $\frac{dw}{dx}=\tan\theta \sim \frac{\lambda}{w} = c_{\lambda}(h/w)^{1/3} $, where the second equality follows from eq. (\ref{lambda}).
Using $w(0) = 0$ we can integrate the previous equation to obtain
\begin{equation} \label{w}
w = c_{\theta}h^{1/4}x^{3/4}.
\end{equation}
This form with $c_{\theta} = 2.4$ is in good agreement with both experiment and simulations (Fig. \ref{fig3}e). Once again, our result is analogous to the case of fully plastic tearing \cite{wierzbicki}, where the above scaling relation arises by postulating that cracks advance perpendicular to tensile stress in a purely kinematic model. Again this is because geometry rules the crack path.

\begin{figure}[t]
\begin{center}
\includegraphics[width = 75mm]{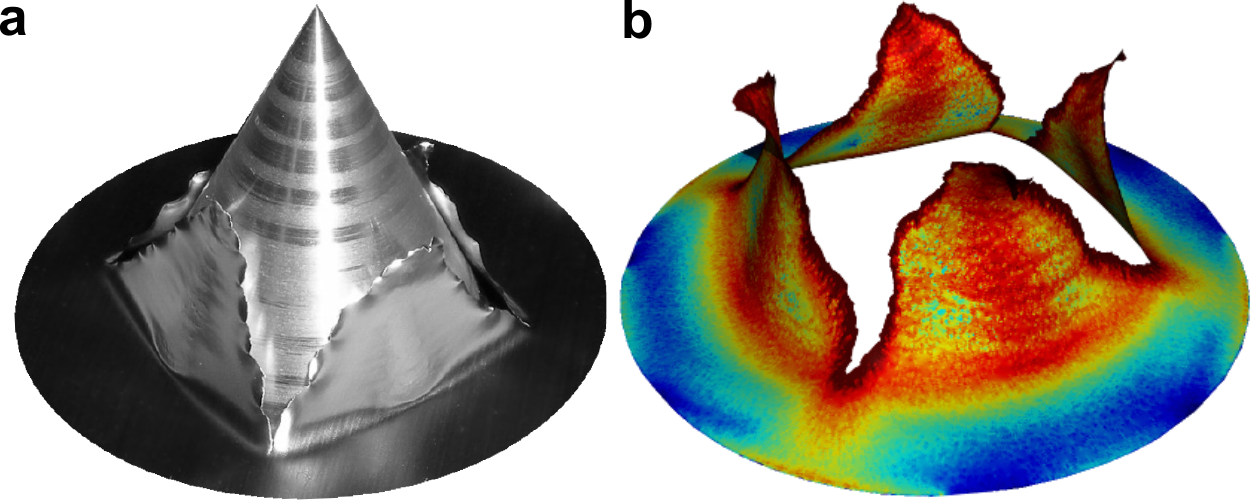}
\end{center}
\caption{ 
(color online). Perforating (a) aluminum film or (b) simulated ductile sheet ($\sigma_y/E = \gamma_s/20$ with $\gamma_s = 0.4$) displays radial cracking with four petals. 
Color indicates deformation energy density (red = high; blue = low).
 }
\label{fig4}
\end{figure}

We conclude with a discussion of tearing induced by the motion of a tool through the sheet in the out-of-plane direction (Fig. \ref{fig4}a). This leads to petal cracks which
has been studied previously in ductile sheets using a combination of experiments and scaling concepts \cite{vermorel, wierzbicki_petalling}.  We complement these results using numerical simulations of tearing driven by a frictionless conical tool. The perforation force  $F = \partial{U_t}/\partial{R}$, where $R$ is the radius of the intersection of the tool and original plane of the sheet and $U_t$ is the total energy of deformation and fracture. As in concertina tearing, cracks are supplied with energy by stretching of the sheet between their tips. Assuming that the amplitude of the deformations is $A$, with $A \ll R$ implying a polygonal hole,  the fracture energy is given by $U_c = n \Gamma h R_c$, where $R_c = R/\cos{\Theta}$ is the distance of the crack tips from the center of the perforating hole, with $\Theta = \pi/n$ the half angle between cracks. In the case of brittle fracture, the resisting force   $F \approx \partial{U_c}/\partial{R} = n \Gamma h / \cos{(\pi/n)}$. The number of cracks can be predicted by minimizing $F$ which occurs for $n = 4$ and yields $F = 5.66 \Gamma h$. Again this result is to be contrasted with  the ductile case where plastic bending of the petals contributes significantly to the force, and yields an optimum number of petals $n=4$ \cite{wierzbicki_petalling} suggesting a universal geometry consistent with our simulations for both brittle and ductile materials (Fig. \ref{fig4}b). 

Our study highlights both the morphological complexity and the geometrical underpinnings of driven cracks in brittle and ductile thin sheets. Using numerical simulations, we have explored the forms that result and highlighted the qualitative differences between the oscillatory tearing in brittle sheets mediated by a single crack, and the concertina folding and tearing mediated by two cracks in ductile materials.  Simple scaling laws allow us to explain our observations, which we also corroborate with experiments. Interestingly, our results are similar to those obtained using purely plastic analyses and highlight the role of geometry again in these systems. A minimal phase diagram characterizes the morphological phase space, and suggests approaches for the controlled tearing of thin films for structure and function.  

\acknowledgments 
We acknowledge the NSF Harvard MRSEC, the Finnish Cultural Foundation (TT) and the MacArthur Foundation (LM) for support, and Z. Suo and J. W. Hutchinson for useful discussions.

\end{document}